\documentclass[aps,prl,superscriptaddress,amssymb,twocolumn,showpacs,a4]{revtex4}

\usepackage{epsfig}
\usepackage{graphics}
\usepackage{graphicx}

\begin{document}


\title{Shearing or Compressing a Soft Glass in 2D: Time-concentration superposition}


\author{Pietro Cicuta}
\affiliation{Cavendish Laboratory, University of Cambridge,
Madingley Road, Cambridge CB3 0HE, U.K.}
\author{Edward J. Stancik}
\affiliation{Department of Chemical Engineering, Stanford
University, Stanford, California 94305-5025}
\author{Gerald G. Fuller}
 \email[]{ggf@stanford.edu}
\affiliation{Department of Chemical Engineering, Stanford
University, Stanford, California 94305-5025}



\begin{abstract}
We report surface shear rheological measurements on  dense
insoluble monolayers of micron sized colloidal spheres at the
oil/water interface and of the protein $\beta$-lactoglobulin at
the air/water surface. As expected, the elastic modulus shows a
changing character in the response, from a viscous liquid towards
an elastic solid as the concentration is increased, and  a change
from elastic to viscous as the  shear frequency is increased.
Surprisingly,  above a critical packing fraction, the complex
elastic modulus curves measured at different concentrations can be
superposed to form a master curve, by rescaling the frequency and
the magnitude of the modulus. This provides a powerful tool for
the extrapolation of the material response function outside the
experimentally accessible frequency range. The results are
discussed in relation to recent experiments on bulk systems, and
indicate that these two dimensional monolayers should be regarded
as being close to a soft glass state.
\end{abstract}


\maketitle

 The rheology of systems with  very large
surface to volume ratios, such as foams or emulsions, depends on
the flow behavior of material that is added to provide stability
and remains constrained to the surfaces. This has  been
particularly investigated in the case of foams \cite{stone99,
weaire00}, highlighting the need for experiments on model systems.
 Insoluble surface films (Langmuir
monolayers) represent the most basic model system, and
quantitative measurements of their viscous and elastic response
when subject to a shear deformation are possible with modern
instruments \cite{fuller99,zasadzinski02}. Understanding flow and
dynamical transitions in this simple geometry promises to aid in
the design of cosmetics and food products with desirable textures,
and to be of value in polymer processing and oil recovery, the
industries that have traditionally driven the interest in this
field.

In this Letter, viscoelasticity is investigated quantitatively
 on two very different  systems that are
confined to fluid interfaces, where flow is effectively two
dimensional. Latex colloids have a non-deformable hard core, and
we show how  a soft solid is formed and the system progressively
jams as the close packing concentration is approached. The
$\beta$-lactoglobulin proteins, which partially unfold at
surfaces, can (like polymer chains) be described as soft spheres.
A surprising scaling behavior, similar to recent results on three
dimensional systems \cite{weitz00}, is reported for both
monolayers. We believe that the generality of behavior between two
(2d) and three (3d) dimensions should be of value and interest for
both the theoretical understanding and the numerical simulations
in this field, and also sheds light onto the origin of monolayer
viscoelasticity and other related glassy behavior such as long
time scale stress relaxation or aging.

In contrast to surface techniques, instrumentation to perform bulk
rheology is well established and widespread, and  the flow
behavior of soft matter bulk systems has been extensively
studied~\cite{larson99}.
There has recently been  much interest in the dynamics of soft
matter systems, from foams~\cite{durian95} to concentrated
emulsions and colloidal suspensions~\cite{weitz95b, weitz95} that
become jammed at high density.
These are materials that respond like elastic solids to small
stresses but they have a yield modulus and, like pastes, flow
above a threshold stress. They are known phenomenologically as
Bingham bodies~\cite{larson99}. In viscoelasticity measurements on
these systems the complex shear modulus $G^*=G'+iG''$ is measured,
and an elastic plateau $G'(\omega)>G''(\omega)$ is typically
observed at low frequency, crossing over to a viscous response
$G''(\omega)>G'(\omega)$ at high frequency. In this context
quantitative  rheological measurements exist only on three
dimensional, bulk systems. In two dimensions it has only recently
been possible to perform a relative measurement of the divergence
of viscosity upon close packing of solid lipid
domains~\cite{zasadzinski02}.

Well established methods exist to perform experiments on
macromolecules that are irreversibly confined at the air/water or
oil/water interfaces in a Langmuir trough: the surface
concentration $\Phi$ is varied by a sweeping barrier, the osmotic
pressure $\Pi$ is determined by measuring the interfacial tension
$\gamma$ and using $\Pi=\gamma_0-\gamma$ where $\gamma_0$ is the
surface tension of a clean interface, the temperature is fixed by
thermostating the subphase liquid and finally the surface can be
easily imaged with optical microscopy.  Nima (Coventry, U.K.) and
KSV (Helsinki, Finland) troughs are used, with either platinum or
filter paper Wilhelmy plates.
   The
   colloid monolayers are made of 3.1$\mu$m diameter Polystyrene
   spheres  (Interfacial Dynamics Corporation), with a surface charge density of 9.1$\mu$C/cm$^2$.
   The colloids  are dispersed onto the interface between an aqueous solution (0.01M NaCl) and
   decane (Fischer). The same methods are followed as in \cite{stancik02}, and monolayers of
   similar particles have
   been investigated in
   \cite{aveyard00a}. The milk protein  $\beta$-lactoglobulin
 is obtained from Sigma (mixture of A and B types, bovine milk, 90\% pure) and used without
  further purification\footnote{In
  the present study a water (MilliQ) solution subphase,  0.01M phosphate
   buffer and 0.1M NaCl having  $p$H6.1 is used, and  the proteins
   are spread    by successively touching the air/liquid surface with $\sim$2$\mu$l drops of a
   concentrated protein solution (1mg/ml in water).}. Monolayers of this protein
  have been studied extensively because of their importance in the
  food industry \cite{mobius98}. Data is collected for about 15~minutes
  at each concentration, and over the timescale of 2~hours
  no aging effect has been noticed.
 Full details of our experiments, comprising creep and stress
relaxation measurements, will be published elsewhere.
\begin{figure}[t]
           \epsfig{file=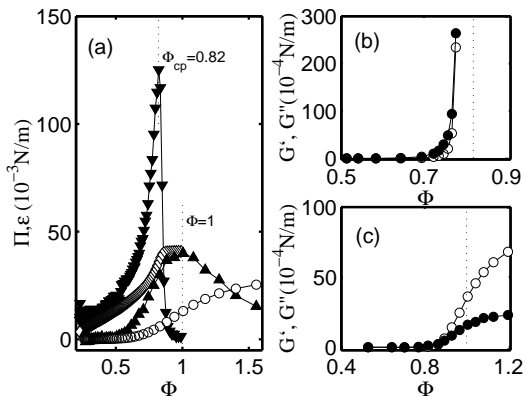,width=7cm}
 \caption{(a)~Two dimensional osmotic pressure $\Pi$ as a function of the surface concentration
 $\Phi$, for ($\diamond$)~3$\mu$m colloids and ($\circ$)~$\beta$-lactoglobulin.
  The  dotted lines correspond to the maxima in the dilational compression modulus
  $\varepsilon$, which occur
  at close packing ($\Phi=0.82$) for the colloid monolayer~($\blacktriangledown$)
  and at $\Phi=1$ for $\beta$-lactoglobulin~($\blacktriangle$). The elastic  ($\circ$)~$G'$  and
  viscous ($\bullet$)~$G''$ components of the shear modulus at $\omega\simeq1$Hz are plotted as a function
of the concentration for~(b)~the colloid  and (c)~the
$\beta$-lactoglobulin monolayers.  \label{fig1}}
\end{figure}

The osmotic pressure dependence on the surface concentration
 is related to the intermolecular interactions. $\Pi-\Phi$
isotherms for the two systems studied are shown in
Figure~\ref{fig1}(a) and are well known in the literature
\cite{mobius98, aveyard00a}. Determining $\Pi$ is the simplest
measurement on a monolayer, and the isotherms serve as a reference
to establish the concentration of the monolayer in separate
experiments. The packing fraction $\Phi$ for the colloids is
determined by optical microscopy, and the osmotic pressure of the
colloidal layer is mainly due to electrostatic repulsion
\cite{aveyard00a}. The maximum in the dilational modulus
$\varepsilon(\Phi)=1/\Phi\,d\Pi/d\Phi$, see Figure~\ref{fig1}(a),
occurs at $\Phi=0.82$, which is the two dimensional random close
packing value for disks. For  $\Phi\gtrsim 0.82$ the colloids are
forced out of the  monolayer plane. It is well known that proteins
can unfold at an interface, and that  the initial upturn region of
their isotherms is well described by a polymer-like semi-dilute
regime scaling law which starts at the overlap concentration of
coils \cite{mobius98}. Within this regime, in 2d the polymer coils
are expected to be segregated and to be progressively compressed.
The protein surface concentration in Figure~\ref{fig1} has been
normalized so that the scaling regime terminates at $\Phi=1$
(assuming no loss to the subphase on spreading, at $\Phi=1$ the
surface concentration of $\beta$-lactoglobulin is 1.45mg/m$^2$).
This concentration corresponds to the peak in the compression
elastic modulus $\varepsilon(\Phi)$. For polymer systems the
maximum  signals
  the transition to a concentrated regime, in which there is still
space for additional  monomers on the surface. The proteins act
like soft disks and the monolayers can be compressed above
$\Phi=1$.

%
%
%
%
%

Rheological measurements on monolayers are performed with an
interfacial stress rheometer (ISR), which has been described in
detail elsewhere \cite{fuller99}.  It consists of a 5cm long and
0.5mm diameter  magnetized rod confined to the plane of the
interface, set into oscillation by applying a sinusoidally
time-dependent magnetic field gradient and tracked by projecting
the rod image through an inverted microscope onto a linear
photodiode array. A half-cylinder glass channel of length 10cm and
radius 3.2mm was kept submerged in the water phase and its
position was adjusted so that the interface meniscus was pinned to
the inside of the glass wall, ensuring well defined and
reproducible open channel flow boundary conditions. After
calibration of the instrumental parameters by performing reference
runs on clean water/air or water/oil interfaces, the amplitude and
phase of the rod motion can be analyzed to give the dynamic
surface shear modulus $G^*(\omega)$:
\begin{eqnarray}
G^*(\omega)\,=\,\frac{\sigma_s}{\gamma_0} \exp\left(i
\delta(\omega)\right), \label{eq1}
\end{eqnarray}
  where $\sigma_s$ is
the amplitude of the applied sinusoidal stress  with frequency
$\omega$, $\gamma_0$ is the amplitude of the resulting strain,
having the same frequency $\omega$ and a phase difference
$\delta(\omega)$. Measurements are performed for fixed strain
amplitude of 3\%, at a range of frequencies. Strain sweep
experiments, not shown, are used to check that the response is in
the linear regime.

The shear modulus $G^*(\omega)$ shown   in Figure~\ref{fig1}(b)
and~(c)
  is determined as the average of
measurements at frequencies between 0.7 and~1Hz, at successive
concentrations.
 For colloidal spheres,  Figure~\ref{fig1}(b), the complex modulus $G^*$  shows an initial
upturn  at $\Phi \simeq 0.64$ and, at the frequency of this
experiment, a predominantly viscous response
$G''(\omega)>G'(\omega)$ is observed across the measured
concentration range.  The  modulus $G^*$ of the
$\beta$-lactoglobulin monolayer, see Figure~\ref{fig1}(c), shows
an upturn at $\Phi=0.77$  and $G''(\omega)>G'(\omega)$ only for
$0.77<\Phi<0.87$. There is
 a wide range of concentrations where the elastic
modulus dominates the loss modulus. In contrast to the colloidal
layer, there is a change in the concavity of $G'$ and $G''$,
occurring at the same concentration ($\Phi\simeq 1$) where the
elastic compression modulus is maximum.

\begin{figure}[t]
           \epsfig{file=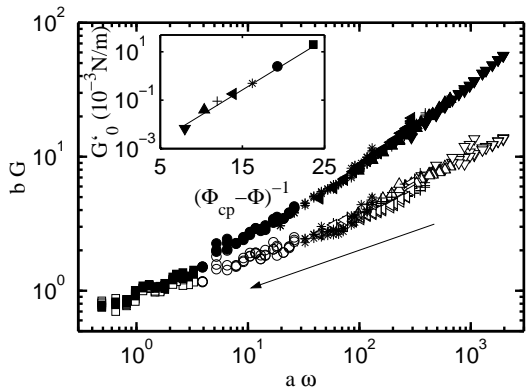,width=7cm}
 \caption{Master curve showing log plots of the scaled values of (open symbols)~$G'$  and (solid symbols)~$G''$
 against the scaled frequency for the
monolayer of colloidal particles. The arrow indicates the
direction of increasing surface concentration, which coincides
with increasing magnitude of moduli.  As described in the text,
the time-concentration superposition enables the
 extrapolation of data outside the experimental
frequency window. The inset shows an exponential divergence of the
elastic modulus as the  concentration approaches the close-packing
fraction $\Phi_{cp}$. \label{figcoll}}
\end{figure}

\begin{figure}[t]
           \epsfig{file=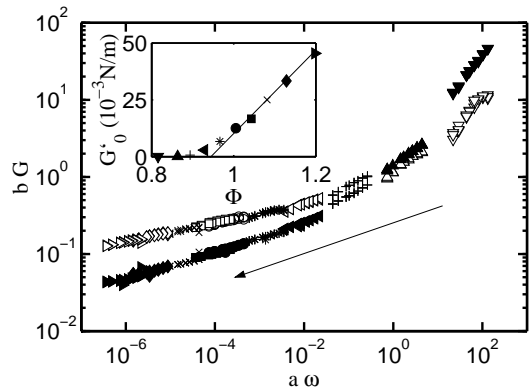,width=7cm}
 \caption{Master curve showing log plots of the scaled values of (open symbols)~$G'$  and (solid symbols)~$G''$
against the scaled frequency for $\beta$-lactoglobulin. The arrow
indicates  increasing concentration.  As in Figure~\ref{figcoll}
the inset presents extrapolated data, which shows a linear
dependence of the elastic modulus on concentration for $\Phi\geq
1$. \label{figblac}}
\end{figure}

Trappe and Weitz \cite{weitz00} recently showed for a 3d system of
very dilute weakly attractive colloidal particles that the
viscoelastic moduli obtained as a function of frequency at
different volume fractions (and even for different interaction
potentials) could be scaled on a single master curve. The overlap
for different volume fractions occurs because of a self-similar
response at different concentrations, and implies the same change
in behavior if the concentration is reduced or the shear rate is
increased.  With the ISR it is possible to measure the shear
elastic moduli of Figure~\ref{fig1}(b) and~(c)  over almost two
decades in frequency, between 0.015 and 1.1Hz,  enabling a study
of the response behavior as a function of frequency (as well as
concentration) and an analysis similar to \cite{weitz00}. In
contrast to the system studied by Trappe and Weitz,  in this work
flow is in two dimensions, the packing is dense,  the interactions
are repulsive and the structure does not appear fractal. We find
that above a threshold concentration $\Phi_c=0.7$ for the colloids
and 0.8 for $\beta$-lactoglobulin, the frequency dependent
measurements of $G^*(\omega)$ taken at different concentrations
can be overlayed on a master curve by rescaling the frequency
$\omega$ by a factor {\it{a}} and $G^*$ by {\it{b}}. The
concentration $\Phi_c$ can be understood to be the cross-over
point at which the system develops a stress-bearing network
\cite{weitz01} and it is expected to depend on both the
temperature and the details of the interaction potential.  In
principle, the concavity of $G^*(\omega)$ is such that the
conditions of continuity and smoothness  determine {\it{a}} and
{\it{b}}, however in practice the  overlap between consecutive
datasets is small and the choice of {\it{a}} and {\it{b}} is made
by hand. The main results of this Letter are Figures~\ref{figcoll}
and~\ref{figblac}, which  show the master curves for  the two
systems.

The colloidal monolayer response,  Figure~\ref{figcoll}, is most
elastic at low frequencies and  predominantly viscous at high
frequency. The rheological behavior of the colloid particles  can
be expected to be close to that of hard disks, and the value of
$\Phi_{c}$ for the colloids coincides with the entropically
induced freezing transition density for mono-disperse hard disks
in 2d \cite{mitus97}. This suggests that the master curve of
 Figure~\ref{figcoll}, which is at present an empirical observation, is the material
response function of a soft 2d solid, in which the motion of each
particle  is hindered by the cage-like structure of its neighbors.
There is a remarkable similarity with the rheological curve for
dense three dimensional colloidal suspensions~\cite{weitz95},
including the  high frequency shear-thinning $G''\sim\omega^{0.5}$
limiting behavior~\cite{shepper93}.   The existence of a master
curve allows the extrapolation of  data to experimentally
inaccessible timescales and moduli, in the same way as allowed by
time-temperature superposition in polymer systems \cite{larson99}.
By this approach it is possible to measure,  at extremely low
concentrations, the value of the modulus where the viscous and
elastic components are equal, $G'_0$. The inset in
Figure~\ref{figcoll} shows that  the modulus diverges
exponentially as the  concentration approaches close packing,
which is the concentration of the glass transition of hard
disks~\cite{santen00}. The same divergence has been recently
reported for the viscosity of 3d colloidal hard-sphere dispersions
approaching the glass transition~\cite{russel02}, and the
exponential dependance was explained in terms of cooperatively
rearranging groups of particles. It should be possible to test
this hypothesis directly  by 2d particle tracking  in future
monolayer experiments.

The master curve for $\beta$-lactoglobulin, Figure~\ref{figblac},
has the same qualitative shape as for the colloidal monolayer, but
the rescaling of data is  most evident at lower frequencies.  The
cross-over to a solid-like  response dominated by the elastic
component of the modulus, which is at the experimental limit of
the measurements on colloids, can be clearly seen here. The
protein layer can be thought of as a system of compressed and
deformed disks. In the limit of high concentration (or low
frequencies), the elastic moduli follow power law frequency
dependencies $G'\sim \omega^{0.1}$ and $G''\sim \omega^{0.2}$.
These very small exponents are consistent with the soft glassy
rheology model \cite{cates97} proposed for materials close to the
glass transition, which does not however account for  the
time-concentration superposition reported here. The master curve
enables the measurement of $G'_0$ at high density and the study of
its concentration dependence, shown in the inset of
Figure~\ref{figblac}. Above $\Phi\simeq 1$ the modulus scales
linearly with the concentration. This is the same behavior
observed in 3d compressed emulsions~\cite{weitz95c} above the
random close packing concentration, and supports the analogy
between the $\beta$-lactoglobulin monolayer and a system of soft
disks.

\begin{figure}[t]
           \epsfig{file=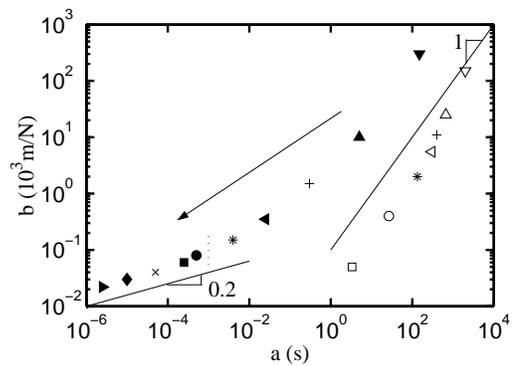,width=6.8cm}
 \caption{Log plot of the scaling factor
{\it{b}} {\it{vs.}} {\it{a}}, for  colloids~(open symbols) and
$\beta$-lactoglobulin~(solid symbols). The symbols correspond to
different concentrations and are consistent with
Figures~\ref{figcoll} and~\ref{figblac}, enabling the experimental
data to be reconstructed. The arrow indicates increasing
concentration, and the dotted line delimits the region  $\Phi\geq
1$ in the $\beta$-lactoglobulin monolayer. For both systems the
scaling factors have been chosen so that the crossover from
elastic to viscous response occurs when the reduced frequency and
modulus are equal to~1. \label{figmaster}}
\end{figure}

 The
relationship between the scaling factors for the frequency and the
modulus contains information on the underlying physics. Trappe and
Weitz~\cite{weitz00} found a linear relationship in their
attractive particle system, and proposed a simple explanation in
which the particle network elasticity increases as a function of
$\Phi$ in agreement with results for elastic percolation, and the
viscosity is due to coupling with the solvent. A consequence of
their model is  that at large frequencies  $G''(\omega)$ tends
asymptotically to $\omega \eta_{0}$, where $\eta_{0}$ is the
solvent viscosity. This does not happen in our data, and we
believe that in the dense systems studied in this work  the
time-concentration superposition has a different origin. For the
colloidal monolayer $a$~and~$b$ are approximately linearly
dependent, as shown in Figure~\ref{figmaster}. This means that if
the shear modulus grows as a function of the concentration, the
dynamics slows down with the same concentration dependance.   This
can be understood trivially for the viscous component $G''$: The
viscosity of the 2d suspension grows as the concentration
 increases (and $G''$ is proportional to the viscosity),
and  the  dynamical timescales are inversely proportional to the
viscosity.
 The scaling factors for
$\beta$-lactoglobulin do not hold the same relationship throughout
the concentration range, meaning that there are likely to be
complex sources  of elasticity and viscosity for the system of
highly compressed soft particles.

  The measurements reported above have
explored the similarity in viscoelastic response between surface
monolayers and bulk systems, and between systems with very
different interaction potentials. They indicate  that models
should be general to both dimensions and might stimulate further
experiments  to probe specific issues in soft glassy materials,
such as the spatial and temporal extent of dynamical events, that
could most easily be realized in surface monolayers.

\begin{acknowledgments}
We thank V.~Trappe, L.~Cipelletti, M.E.~Cates and P.G.~Olmsted for
very useful comments and discussions.
\end{acknowledgments}

\bibliography{pstanford}

\end{document}